\documentclass[prl,twocolumn,showpacs,
preprintnumbers,amsmath,amssymb,showkeys]{revtex4}
\usepackage{epsfig}
\usepackage{graphics}
\usepackage{color}
\usepackage{amsmath}

\newcommand{\etal}[1]{{\it et al.}}

\newcommand{\HHCO}{H$_2$CO}

\newcommand{\DDO}{D$_2$O}
\newcommand{\HHO}{H$_2$O}
\newcommand{\NDDD}{ND$_3$}

\newcommand{\cminv}{cm$^{-1}$}

\newcommand{\ket}[1]{\ensuremath{\left|#1\right\rangle}}

\begin{document}

\title{Water vapor at a translational temperature of one kelvin.}
\author{T. Rieger}
\author{T. Junglen}
\affiliation{Max-Planck-Institut f\"ur Quantenoptik,
Hans-Kopfermann-Str. 1, D-85748 Garching, Germany}
\author{S.A. Rangwala}
\altaffiliation{Raman Research Institute, C. V. Raman Avenue,
Sadashivanagar, Bangalore 560080, India}
\author{G. Rempe}
\author{P.W.H. Pinkse}
\affiliation{Max-Planck-Institut f\"ur Quantenoptik,
Hans-Kopfermann-Str. 1, D-85748 Garching, Germany}
\author{J. Bulthuis}
\affiliation{Department of Physical Chemistry and Laser Centre,
Vrije Universiteit, De Boelelaan 1083, 1081 HV Amsterdam, The
Netherlands}
\date{\today, PREPRINT}

\begin{abstract}

We report the creation of a confined slow beam of heavy-water
(D$_2$O) molecules with a translational temperature around 1 kelvin.
This is achieved by filtering slow D$_2$O from a thermal ensemble
with inhomogeneous static electric fields exploiting the quadratic
Stark shift of D$_2$O. All previous demonstrations of electric field
manipulation of cold dipolar molecules rely on a predominantly
linear Stark shift. Further, on the basis of elementary molecular
properties and our filtering technique we argue that our D$_2$O beam
contains molecules in only a few ro-vibrational states.


\end{abstract}

\maketitle

\vspace{1cm}


Cold dilute molecular systems are rapidly emerging as a front line
area at the interface of quantum optics and condensed matter
physics~\cite{EPJD04}. An increasing subset of this activity centers
around the creation of cold dilute gases of molecules possessing
electric dipole moments. These in particular, owing to their
long-range anisotropic interaction, hold the promise of novel
physics, where two- and many-body quantum properties can be
systematically studied. Cold dilute gases of dipolar molecules can
be produced by forging a tight bond between two chemically distinct
species of laser-cooled atoms, e.g. RbCs~\cite{SagePRL05}.
Alternatively, cold dilute gas ensembles can be created by
buffer-gas loading~\cite{Weinstein98} or electric-field manipulation
of naturally occurring molecules like
\NDDD~\cite{BethlemNature00,JunglenEPJD04}, \HHCO~\cite{Rangwala03},
metastables like CO~\cite{BethlemPRL99} or radicals like
YbF~\cite{SauerPRL04}, OH~\cite{BochinskiPRL03},
NH~\cite{MeerakkerPRA01}. So far all the cold molecules made
available with electric-field-based methods have a Stark effect (in
their relevant states) which is predominantly linear in the
important range up to 150\,kV/cm.

Here we report the creation of a slow beam of heavy-water (\DDO)
molecules, which experience a quadratic Stark effect. The cold \DDO\
molecules are filtered from a room-temperature thermal
gas~\cite{Rangwala03} and have a translational temperature around 1
kelvin. Because the Stark shifts are quadratic in the electric
field, it follows that forces exerted by inhomogeneous electric
fields are relatively small for \DDO\ compared to molecules with
similar dipole moments but with linear Stark shifts. It is therefore
by no means obvious that significant quantities of slow \DDO\
molecules can be produced by means of electric-field-based methods.
Our experimental result therefore underlines the versatility of the
velocity-filtering method. It is an enabling step towards future
trapping of molecules for which the ratio of elastic to inelastic
collisions is expected to be more favorable than for molecules with
linear Stark shifts~\cite{BohnPRA01}. An additional advantage of the
quadratically Stark-shifted molecules like \DDO\ is the possibility
to perform precise spectroscopic measurements insensitive to stray
electric fields, to the first order. Moreover, water is abundant in
interstellar space at low densities and temperatures from a few
kelvin upward, playing an important role in the chemistry of
molecular clouds~\cite{Spaans01}. The conditions in these clouds are
remarkably close to those achieved in our experiment, opening up the
possibility to investigate in the laboratory chemical reactions
under conditions found in space.

This Letter is structured as follows. First we discuss general
features of Stark shifts of molecular states with particular
references to \DDO. We then present our experimental work with \DDO.
This is followed by arguing from first principles that the resulting
beam of \DDO\ is dominated by only 4 rotational states, despite
starting with a thermal source of molecules at 300 K.

Several techniques have recently been developed to manipulate
molecules~\cite{EPJD04} with electric fields. All of these exploit
the Stark effect to exert a force on the molecules. In contrast to
atoms, molecules can have a permanent electric dipole moment. Such
molecules have much larger Stark shifts than non-polar molecules.
However, as described below, a large dipole moment alone is not
enough to have a strong Stark effect. The direction and magnitude of
the force exerted on the molecule in an inhomogeneous electric field
depends on the details of the molecular rotational state. Assuming
the Stark shift to be a monotonic function of the electric field,
the molecule can be either in low-field-seeking (LFS) or
high-field-seeking (HFS) states, depending on the sign of the Stark
shift.

The condition for having a linear Stark effect is that the component
of the dipole moment $\vec{d}$ along a space-fixed direction, we can
choose $\vec{z}$, is non-vanishing (i.e. $\langle \vec{d}\cdot
\vec{z}\rangle \neq 0$). Strictly speaking this requires a finite
electric field, but it can be arbitrarily low. The linear Stark
shift is typically found in symmetric top molecules and is
proportional to $|\vec{d}| K M$, with $K$ and $M$ representing the
projection of the total angular momentum $\vec{J}$ on the molecular
symmetry axis and on the $z$-axis, respectively. Of course, no
first-order Stark effect occurs if either $M$ or $K$ -- or both --
are zero. If the degeneracy in zero field is lifted, e.g. by
fine-structure splitting, inversion doubling, nuclear quadrupole
interaction in a symmetric-top molecule or $\Lambda$-doubling in
linear molecules, the Stark splitting will no longer be linear in
the limit of zero field. However, often those interactions are small
enough as to lead to a nearly linear Stark splitting in the applied
electric field range.

In general, the linear Stark effect condition, $\langle \vec{d}\cdot
\vec{z} \rangle \neq 0$, is not fulfilled in asymmetric top
molecules. Under certain conditions, however, polar asymmetric
molecules can also exhibit (nearly) linear Stark shifts. If the
asymmetry is weak, the states that correspond to $K \neq 0 $ in the
prolate or oblate top limit, will always be close to being
degenerate. Those states show nearly linear Stark shifts if they are
coupled by the Stark interaction. This is the case if the dipole is
along the $a$-axis in the prolate limit, or along the $c$-axis in
the oblate limit, where we follow the convention to label the axis
with the smallest moment of inertia and hence the largest rotational
constant with $a$, the intermediate axis $b$ and the axis with the
largest moment of inertia $c$. An example is the nearly symmetric
(prolate) top molecule \HHCO.

\begin{figure}[htb]\begin{center}

\epsfig{file=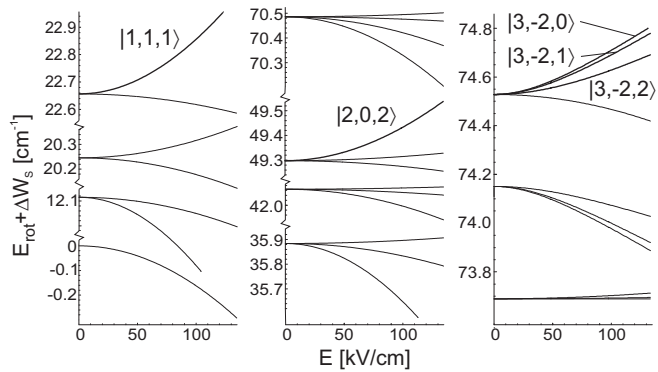,width=0.48 \textwidth}%
\caption{The lowest rotational energies of \DDO\ as a function of
the applied electric field $E$. The 5 most abundant states
$|J,\tau,M>$ in the guided beam are indicated. The Stark shifts are
obtained by numerically diagonalizing the Stark Hamiltonian
($J=0\dots 12$), following Ref.~\cite{HainJCP99}.}
\label{WaterLevels}
\end{center}\end{figure}
True asymmetric top molecules in general have quadratic Stark
shifts. Exceptions can occur for some states, if the dipole is
oriented along the axis of largest or smallest moment of inertia.
For molecules with their dipole oriented along the $b$-axis, we
found no exceptions: all rotational states have a non-linear Stark
shift. Water, both \HHO\ and \DDO, presents such a case and a few
levels of \DDO\ are depicted in Fig.~\ref{WaterLevels}. The
quadratic behavior is obvious; only for the highest most abundant
states, the $|J=3, \tau, M\rangle$ states (where $\tau$ is a pseudo
quantum number labelling the state), a deviation is found. Moreover,
the large rotational constants~\cite{SchawlowTownes} for \DDO,
$A=15.394\,$\cminv, $B=7.2630\,$\cminv, $C=4.8520\,$\cminv, imply
large rotational level spacings. Avoided level crossings are neither
expected nor found, and second-order perturbation theory is a
reasonable approximation for the Stark shift computation of \HHO\
and \DDO. Moreover, since the contribution to the perturbation from
each coupled pair of states is inversely proportional to the energy
gap between the pair, the shift will be proportional to the density
of (rotational) states, and therefore very small for the sparse
rotational spectrum of \DDO\ and \HHO. Our choice of working with
\DDO\ as opposed to \HHO\ has partly to do with the larger Stark
shifts of \DDO\ because of its smaller rotational constants. The
treatment of the general case of an asymmetric molecule, where the
dipole is not necessarily along one of the principal axes, is, of
course, more involved.

\begin{figure}[htb]\begin{center}
\epsfig{file=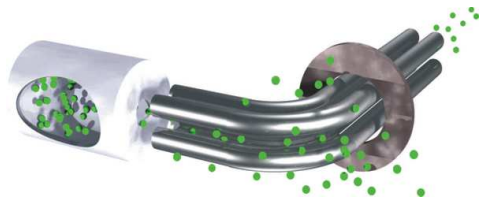,width=0.35\textwidth}%
\caption{(Color online) Schematic of the experiment. On the left
is the effusive source, which injects thermal \DDO\ molecules
into the 4-wire guide. Neighboring electrodes have opposite
polarity, creating a quadrupolar electric field. Molecules that
are slow enough are guided through the first and second (not
shown) $90^{\rm o}$ bends and are finally detected by a mass
spectrometer.}
\label{idea}
\end{center}\end{figure}

Our apparatus~\cite{JunglenEPJD04} is depicted in Fig.~\ref{idea}.
It consists of a room-temperature effusive thermal source, which
injects \DDO\ molecules directly between four 50\,cm long electrodes
set up in a quadrupole arrangement, with neighboring electrodes
having opposite polarities. The guide has two $90^{\rm o}$ bends
with a radius of curvature of 25\,mm. The quadrupolar electric field
defines a two-dimensional potential well. This well has a depth that
depends on the internal molecular state, e.g. for the $\ket{J,\tau,
M}=\ket{1,1,1}$ state with a positive Stark shift of 0.20\,\cminv\
at 100\,kV/cm, the depth amounts to 0.29\,K. Molecules with
transverse kinetic energy exceeding the potential depth escape the
guide. In the bends the longitudinally fast molecules escape while
the slow ones are kept due to the action of the centripetal force.
These are guided through two differential pumping regions into an
ultrahigh vacuum chamber for mass-spectrometric detection at the end
of the electrodes. Heavy water is convenient for this purpose, as
the background at its mass is virtually zero. The longitudinal
velocity distribution of the guided \DDO\ beam was determined by a
time-of-flight method~\cite{Rangwala03} at an escape field, $E$, of
115\,kV/cm. We found a most-probable velocity of 24\,m/s in the
laboratory frame, corresponding to a longitudinal temperature of
$\approx1.4\,$K. The transverse temperature is expected to be on the
order of 0.1\,K, as the guide presents a smaller transverse velocity
cutoff value than the corresponding longitudinal velocity cutoff.

The flux dependence on the escape field, $E$, and hence on the
applied electrode voltage $V$, is characteristic of the nature of
the guided molecules' Stark shift. This can be seen as follows: Let
$x,y$ be directions orthogonal to and $z$ be parallel to the
quadrupolar axis. Let $f_{v_{x,y,z}}$ be functions proportional to
the flux crossing the planes of unit area perpendicular to the
$x,y,z$ axes, respectively~\cite{Pinkse03}. Then $f_{v_{x,y}}\propto
\exp(-v_{x,y}^2/\alpha^{2})$ is bi-directional and $f_{v_z}\propto
v_z \exp(-v_z^2/\alpha^{2})$ is uni-directional along the positive
$z$ axis. Here, $\alpha=\sqrt{2 k_B T/m}$, $k_{\rm B}$ the Boltzmann
constant, $T$ the temperature of the reservoir where the beam
originates from and $m$ the molecular mass. Hence the total guided
flux $ \Phi \propto \int_0^{v_{x,{\rm max}}} d v_x \int_0^{v_{y,{\rm
max}}} d v_y \int_0^{v_{z,{\rm max}}} d v_z f_{v_x} f_{v_y}
f_{v_z}$, where $v_{x,y,z,{\rm max}}$ are the maximal guided
velocities in each direction. As $\alpha \gg v_{x,y,z,{\rm max}}$, $
\Phi \propto v_{x,{\rm max}}\, v_{y,{\rm max}}\, v_{z,{\rm max}}^2$.
The maximum kinetic energy, $U_{\rm k, max}$ is given by the escape
energy of the guide, i.e., the Stark shift in $E$, which is
proportional to the applied electrode voltage, $V$. Hence for
molecules with a linear Stark shift, $\Phi \propto U_{\rm k}^2
\propto V^2$. For molecules with a quadratic Stark shift, $U_{\rm
k}\propto V^2$ and hence, $\Phi \propto U_{\rm k}^2 \propto V^4$.

\begin{figure}[htb]\begin{center}
\epsfig{file=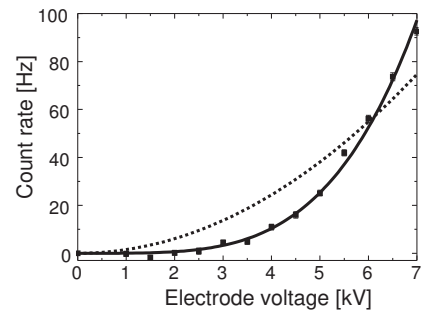,width=0.3 \textwidth}%
\caption{Detector signal versus the electrode voltage $V$. The data
follow a quartic law in $V$, as illustrated by the $V^4$ fit (solid
line). The dotted line shows a $V^2$ fit attempt.}
\label{QuarticPlot}
\end{center}\end{figure}

Our detector, a quadrupole mass spectrometer, converts molecules to
ions by electron-impact ionization. The ions are then mass-selected.
Measurement on various molecules with a linear Stark shift indicate
that the signal is to a good approximation proportional to the
guided flux~\cite{Rangwala03,JunglenEPJD04}. Scaling (for detector
counting efficiencies, angular divergence of the beam exiting the
guide) and corrections (velocity dependent detection, branching
ratios of ionization) are needed to convert our measured count rates
(plotted in Fig.~\ref{QuarticPlot}) to the absolute flux
$\approx7\times10^7\,$s$^{-1}$ at $V=7\,$kV, corresponding to an
electric field depth of the guide of $E=134\,$kV/cm. The error
margin of the flux is estimated to be of the order of a factor 2.
The quartic dependence on $E$ is clearly visible in
Fig.~\ref{QuarticPlot} and proofs the quadratic Stark shift of the
guided molecules. Indeed, with the same apparatus it has been
observed~\cite{Rangwala03,JunglenEPJD04} that for \HHCO\ and \NDDD
(linear Stark molecules), the flux depends quadratically on $V$.

\begin{figure}[htb]\begin{center}
\epsfig{file=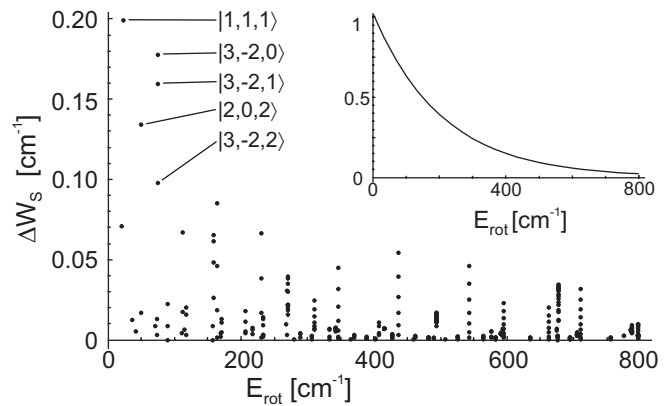, width=0.48 \textwidth}%
\caption{Stark shifts, $\Delta W_{\rm S}$, in an electric field
of 100\,kV/cm of LFS rotational states of \DDO\ in its
vibrational and electronic ground state, as a function of the
zero-field rotational energy, $E_{\rm rot}$. The inset shows the
Boltzmann factor at T=300\,K. The five most Stark-shifted states
are labeled $\ket{J,\tau,M}$. Note that the spin statistical
weighting is not shown.} \label{StarkshiftPlot}
\end{center}\end{figure}

As our slow beam originates from a room-temperature source, many
rotational states are populated. This is illustrated in
Fig.~\ref{StarkshiftPlot}, where the Stark shifts of \DDO\ in a
field of 100\,kV/cm have been plotted as a function of the
zero-field rotational energy. The Stark shifts are obtained by
numerically diagonalizing the Stark Hamiltonian for a rigid
asymmetric rotor, following the procedure of Ref.~\cite{HainJCP99}.
It is known that the rigid-rotor assumption is only a coarse
approximation when estimating the absolute energies of \DDO\ states.
In the present case, however, the approximation is expected to be
good if one is only interested in the Stark shifts and not in the
absolute energies, because the Stark shifts are caused by the
coupling of adjacent states that do not differ much in their sets of
rotational quantum numbers, leading to relatively small sensitivity
to the centrifugal distortion. We have also neglected hyperfine
couplings, which is completely justified in the range of field
strengths used in our experiments. As input for the Stark shift
calculations we took the rotational constants and the dipole moment
$\mu=1.87\,$Debye, which is directed along the
$b$-axis~\cite{SchawlowTownes}. Note that as a general trend, the
Stark shifts decrease with rotational energy.

As discussed earlier the room temperature rotational spectrum of
\DDO\ is very sparse. The populations in the thermal gas in the
effusive source can be estimated by weighting all thermally
populated states with the Boltzmann factor, their orientational
($M$) degeneracy and the spin-statistical weight~\cite{Herzberg}.
Further, only very few of these levels have large enough Stark
shifts to be guided. This selection is much more pronounced for
molecules with a quadratic Stark shift than for molecules with a
linear Stark shift. The intuitive reason is that for molecules with
a quadratic Stark shift, the electric field must first orient the
dipole in space, which is harder for faster rotating molecules.
Indeed, for these molecules the Stark shift~\cite{SchawlowTownes} is
approximately proportional to $ 1/(J+1)$. Hence, the (maximum) Stark
shifts decreases with $J$, as can be seen for \DDO\ in
Fig.~\ref{StarkshiftPlot}. This should be compared with the
dependence of the Stark shift on $J$ of molecules with a linear
Stark shift. For the generic example of a symmetric top, this shift
is $\propto |\vec{d}| K M E /(J(J+1))$ \cite{SchawlowTownes}. As
$(K,M)\in\{-J,\dots,0,\dots,J\}$, the Stark shift of the maximum
($K,M$) for molecules with a linear Stark shift will not decrease
with $J$. In fact, knowing that the flux of \DDO\ molecules is
proportional to the square of the Stark shift, and assuming that
this dependence holds for each state, the four most populated states
contribute more than 70\% of our guided flux. The partial
contributions, zero-field energy and Stark shifts of the 5 most
abundant states are summarized in table~\ref{thetable}. One should
note that the beam purity~\cite{purity} is independent of voltage
changes as long as all the states are quadratic in nature.

\begin{table}
\begin{tabular}{cccc}
State & Contribution [ \% ] & $E_{\rm rot}$ [\cminv] & $\Delta W_{\rm S}$ [\cminv] \\
\hline%
$\ket{3,-\!2,1}$ & 21 & 74.53 & 0.16\\
$\ket{1,1,1}$ & 21 & 22.66 & 0.20\\
$\ket{2,0,2}$ & 17 & 49.30 & 0.13\\
$\ket{3,-\!2,0}$ & 13 & 74.53 & 0.18\\
$\ket{3,-\!2,2}$ & 8 & 74.53 & 0.10 %

\end{tabular}

\caption{The most dominant rotational states $\ket{J,\tau,M}$ of
\DDO\ with their partial contribution to the flux, their zero-field
energy and the Stark shift $\Delta W_{\rm S}$ at
$E=100\,$kV/cm.}%
\label{thetable}
\end{table}

In conclusion, we have demonstrated the effective Stark manipulation
of a polar molecule with quadratic Stark shifts over the range of
applied fields of 0-135\,kV/cm. This experimentally shows the
feasibility of the velocity-filtering method for
quadratically Stark-shifted molecular states. Using this method we
have created water vapor (\DDO) at a translational temperature of
$\approx 1$\,kelvin. Its quadratic Stark effect combined with a
large rotational spacing make \DDO\ a promising molecule for
electric trapping~\cite{RiegerPRL05} and even evaporative cooling.


\begin{acknowledgments}
We acknowledge financial support by the Deutsche
Forschungsgemeinschaft (SPP 1116).
\end{acknowledgments}



\begin{thebibliography}{}

\bibitem{EPJD04}
See, e.g., the special issue on ultracold polar molecules [Eur.
Phys. J. D {\bf 31}, 149 (2004)].

\bibitem{SagePRL05}
J. M. Sage, S. Sainis, T. Bergeman, and D. DeMille, Phys. Rev. Lett.
{\bf 94}, 203001 (2005).

\bibitem{Weinstein98}
J.D. Weinstein, R. deCarvalho, T. Guillet, B. Friedrich, and J.M.
Doyle, Nature (London) {\bf 395}, 148 (1998).

\bibitem{JunglenEPJD04}
T. Junglen, T. Rieger, S.A. Rangwala, P.W.H. Pinkse, and G. Rempe,
Eur. Phys. J. D {\bf 31}, 365 (2004).

\bibitem{BethlemNature00}
H.L. Bethlem, G. Berden, F.M.H. Crompvoets, R.T. Jongma, A.J.A.
van Roij, and G. Meijer, Nature (London) {\bf 406}, 491 (2000).

\bibitem{Rangwala03}
S.A. Rangwala, T. Junglen, T. Rieger, P.W.H. Pinkse, and G. Rempe,
Phys. Rev. A {\bf {67}}, 043406 (2003).

\bibitem{BethlemPRL99}
H.L. Bethlem, G. Berden, and G. Meijer, Phys. Rev. Lett. {\bf 83},
1558 (1999).

\bibitem{SauerPRL04} M.R.
Tarbutt, H.L. Bethlem, J.J. Hudson, V.L. Ryabov, V.A. Ryzhov, B.E.
Sauer, G. Meijer, and E.A. Hinds, Phys. Rev. Lett. {\bf 92},
173002 (2004).

\bibitem{BochinskiPRL03}
J.R. Bochinski, E.R. Hudson, H.J. Lewandowski, G Meijer, and J.
Ye, Phys. Rev. Lett. {\bf 91}, 243001 (2003).

\bibitem{MeerakkerPRA01}
S.Y.T. van de Meerakker, R.T. Jongma, H.L. Bethlem, and G. Meijer,
Phys. Rev. A {\bf 64}, 041401(R) (2001)

\bibitem{BohnPRA01}
J.L. Bohn, Phys. Rev. A {\bf 63}, 052714 (2001).

\bibitem{Spaans01}
M. Spaans and E. van Dishoeck, Astrophys J. {\bf 548}, L217 (2001).

\bibitem{HainJCP99}
T.D. Hain, R.M. Moision, and T.J. Curtiss, J. Chem. Phys. {\bf 111},
6797 (1999).

\bibitem{SchawlowTownes}
C.H. Townes and A.L. Schawlow, {\it Microwave Spectroscopy},
(Dover Publications, Inc., New York, 1975).

\bibitem{Pinkse03}
P.W.H. Pinkse, T. Junglen, T. Rieger, S.A. Rangwala, and G. Rempe,
in {\it Interactions in Ultracold Gases}, M. Weidemü\"uller, C.
Zimmermann (Eds). Wiley-VCH, Weinheim, 2003.

\bibitem{Herzberg}
G. Herzberg, {\it Molecular Spectra and Molecular Structure, I. and
III.} (Van Nostrand Reinhold, New York 1966).

\bibitem{purity}
Note that the guided beam is not in internal equilibrium. The
average rotational energy of the beam corresponds to that of a
thermal gas of 81\,K. The purity, however, is equivalent to that
of a thermal gas of 23\,K, as calculated from the entropy of the
internal distribution.

\bibitem{RiegerPRL05}
T. Rieger, T. Junglen, S.A. Rangwala, P.W.H. Pinkse, G. Rempe,
Phys. Rev. Lett. {\bf 95}, 173002 (2005).

\end{thebibliography}
\end{document}